\documentclass[aps,twocolumn,groupedaddress]{revtex4-2}
\usepackage{graphicx}
\usepackage{empheq}
\usepackage{capt-of}
\usepackage[absolute,overlay]{textpos}
\usepackage{pgf}
\usepackage{fancybox}
\usepackage{xcolor}
\usepackage{xspace}
\usepackage{relsize}
\usepackage{multirow}
\usepackage{verbatim} 
\usepackage[export]{adjustbox}
\usepackage{tcolorbox}
\usepackage{graphicx}
\usepackage{slashed}

\usepackage{color}

\usepackage[T1]{fontenc}

\usepackage{graphicx}
\usepackage{bm,bbm}
\usepackage{amsmath,amssymb}%
\usepackage{amsfonts} 
\usepackage{epstopdf}
\usepackage{color}
\usepackage{isotope}
\usepackage{bbding}

\usepackage[capitalize]{cleveref}

\usepackage{multirow,booktabs,dcolumn}
\renewcommand*{\arraystretch}{1.3}

\usepackage[normalem]{ulem}
\usepackage{tikz} 
\usetikzlibrary{shapes,arrows,tikzmark,positioning,matrix,chains}
  \everymath{\displaystyle}
    \tikzstyle{mathbox} = [inner sep=0pt, anchor=base]
    \tikzstyle{every picture}+=[remember picture]
    \usetikzlibrary{calc}
    \usetikzlibrary{positioning}

\newcommand{\be}{\begin{equation}}
\newcommand{\ee}{\end{equation}}

\newcommand{\bea}{\begin{eqnarray}}
\newcommand{\eea}{\end{eqnarray}}

\newcommand{\bs}[1]{\ensuremath{\boldsymbol{#1}}}

\newcommand{\bit}{\begin{itemize}}
\newcommand{\eit}{\end{itemize}}

\newcommand{\bfi}{\begin{figure}}
\newcommand{\efi}{\end{figure}}

\newcommand{\ben}{\begin{enumerate}}
\newcommand{\een}{\end{enumerate}}
\newcommand{\bbk}{\begin{block}}
\newcommand{\ebk}{\end{block}}
\newcommand{\bex}{\begin{example}}
\newcommand{\eex}{\end{example}}
\newcommand{\bal}{\begin{alertblock}}
\newcommand{\eal}{\end{alertblock}}

\newcommand{\rvec}{\bs{r}}

\newcommand{\nopieft}{\mbox{$\slashed{\pi}$EFT~}}

\newcount\colveccount
\newcommand*\colvec[1]{
        \global\colveccount#1
        \begin{pmatrix}
        \colvecnext
}
\def\colvecnext#1{
        #1
        \global\advance\colveccount-1
        \ifnum\colveccount>0
                \\
                \expandafter\colvecnext
        \else
                \end{pmatrix}
        \fi
}

\newcommand{\Lag}{\mathcal{L}}

\graphicspath{{/}{plots/}}
\begin{document}
\title{The Continuum Spectrum  of Hypernuclear Trios}

\author{M. Sch\"{a}fer}
\affiliation{Czech Technical University in Prague, 
              Faculty of Nuclear Sciences and Physical Engineering, 
              B\v{r}ehov\'{a} 7, 11519 Prague 1, Czech Republic}

\affiliation{Nuclear Physics Institute of the Czech Academy of Sciences, 
              25069 \v{R}e\v{z}, Czech Republic}

\author{B. Bazak}
\affiliation{The Racah Institute of Physics,The Hebrew University, 
               Jerusalem 9190401, Israel}

\author{N. Barnea}
\affiliation{The Racah Institute of Physics,The Hebrew University, 
               Jerusalem 9190401, Israel}

\author{J. Mare\v{s}}
\affiliation{Nuclear Physics Institute of the Czech Academy of Sciences, 
              25069 \v{R}e\v{z}, Czech Republic}


\begin{abstract}
The spectrum of hypernuclear trios composed of a $\Lambda$ baryon and
two nucleons is the subject of an ongoing experimental campaign, aiming to
study the interaction of the $\Lambda$ particle with a neutron, and the 3-body
$\Lambda$-nucleon-nucleon force. 
In this manuscript we utilize baryonic effective field theory at leading order,
constrained to reproduce the available low energy light hypernuclear data,
to study the continuum spectrum of such hypernuclear trios. 
Using the complex scaling method and the inverse analytic continuation in the
coupling constant method we find the existence of a virtual state in the $\Lambda n p$
$J^{\pi}=\frac{3}{2}^{+}$ channel, leading to cross-section enhancement near threshold.
For the $\Lambda n n$ $J^{\pi}=\frac{1}{2}^{+}$ channel we predict a resonance state.
Depending, however, on the value of
the $\Lambda N$ scattering length, the resonance pole moves from the 
physical to the unphysical complex energy sheet within the experimental bounds.
\end{abstract}


\maketitle
\section{Introduction}

Understanding the interaction between nucleons and a $\Lambda$ hyperon is the subject 
of an ongoing experimental and theoretical campaign \cite{GalHunMil16}. 
In the last few years much effort
is dedicated to the study of hypernuclear trios ($\Lambda N N$) aiming to determine the
unknown $\Lambda$-neutron ($\Lambda n$) interaction, and the $\Lambda N N$ 3-body force.
The latter is known to have a crucial effect in the nuclear equation of state 
at high density, and therefore on our understanding of neutron stars.  

The $\Lambda$-nucleon interaction is not strong enough to bind a 
$\Lambda N$ pair, making 
the hypertriton $^3_\Lambda\text{H}(I=0,J^\pi=1/2^+)$
the lightest hypernuclei.
It is weakly bound with a $\Lambda$ separation energy 
$B_{\Lambda} = 0.13 \pm 0.05$~MeV \cite{Davis05}.
The experimental search for other bound hypernuclear
trios has found no evidence  
for the hypertriton state $^3_\Lambda\text{H}^*$, $^3_\Lambda\text{H}(I=0,J^\pi=3/2^+)$, 
indicating that the singlet $s=0$ $\Lambda N$ interaction is somewhat 
stronger than the triplet $s=1$ interaction. 

Recently, the HypHI collaboration \cite{HypHI13} has claimed evidence for
a bound $\Lambda n n$ state, $^3_{\Lambda}n(I=1,J^\pi=1/2^+)$.
However, this observation contradicts theoretical analyses
demonstrating that such a bound state cannot exist.
Since the first calculation by Dalitz and Downs \cite{DD59}, numerous theoretical 
studies of $I=0, 1$ and $J=1/2, 3/2$ $\Lambda NN$ states have been performed, 
confirming the observation that no bound $\Lambda nn$ and 
$^3_{\Lambda}\text{H}(I=0,J^\pi=3/2^+)$
 exist within Faddeev calculations for separable potentials 
\cite{ch1:cit5,ch1:cit6}, chiral constituent quark model of $YN$ interactions 
\cite{GarFerVal07,GarVal14} or the Nijmegen hyperon-nucleon potentials \cite{ch1:cit8}. 
The same conclusion was drawn in \cite{ch1:cit9} within variational calculations using 
$YN$ model, simulating the realistic Nijmegen interaction. 
The $\Lambda nn$ system was also studied within a baryonic (pionless) effective field theory (\nopieft) \cite{Ando15,HH19}, 
however, due to uncertainty in fixing the three-body $\Lambda nn$ force no firm 
predictions of its stability could be made. 

In spite of the theoretical 
consensus regarding a bound $\Lambda nn$, the nature of hypernuclear 
$\Lambda NN$ trios remains a subject of an ongoing 
discussion \cite{HypHI19}. Specifically, the search for the $\Lambda nn$ system 
is a goal of the JLab E12-17-003 experiment \cite{JlabE17003}, and the study of the 
$^3_\Lambda\text{H}(I=0,J^\pi=3/2^+)$ state is part of the JLab proposal 
P12-19-002 \cite{JlabP19002}.

Regardless the apparent interest, the possible existence of $\Lambda nn$ and 
$^3_\Lambda\text{H}^*$ hypernuclear continuum states has been directly addressed 
in only few theoretical works. Calculating zeros of the three-body 
Jost function, Belyaev et al. found a very wide, near-threshold, $\rm \Lambda nn$ 
resonance \cite{BelRakSab08}. Afnan and Gibson \cite{AfnGib15} using Faddeev calculation
and separable potentials, fitted to reproduce $\Lambda N$ and $NN$ scattering
length and effective range, concluded that the $\Lambda nn$ state 
exists as a sub-threshold resonance. They also found that
a small increase of the $\Lambda N$ 
interaction strength shifts the resonance position above threshold and thus yields 
an observable resonance. We are not aware of any direct calculation of the 
$^3_\Lambda\text{H}^*$ continuum state, however, as Garcilazo et al. concluded, there is a hint of near-threshold pole which gives rise to large $\Lambda d$ scattering length in $S=3/2$ channel \cite{GarFerVal07}.  

The aforementioned continuum studies \cite{BelRakSab08,AfnGib15,GarFerVal07,GarVal14}
were limited to $A=3$ systems. Therefore, the predictive power of their interaction 
models was not verified against the available experimental $B_\Lambda$ data in
e.g. 4-body or 5-body $s$-shell hypernuclei. 
In fact, applying a gaussian potential mimicking the low energy behavior 
of the separable potential
of \cite{AfnGib15} we find substantial overbinding in these systems.
Given the relatively poorly known $\Lambda N$ scattering parameters, and the precise
$B_\Lambda$ data, such comprehensive study is called for. 

Motivated by the debate regarding the nature of the hypernuclear 3-body states,
 and the soon to be published JLab E12-17-003 $\Lambda nn$ results \cite{JlabE17003},
in the present work we report on precise few-body 
calculations of the hypernuclear $\Lambda NN$ bound and continuum spectrum, using Hamiltonians constructed 
at leading order (LO) in \nopieft \cite{ConBarGal18}. 
This \nopieft is an extension, including $\Lambda$ hyperons, 
of the $n,p$ nuclear \nopieft~Hamiltonian, first reported in
~\cite{Kol99,BHK00} and more recently used to study lattice-nuclei 
in ~\cite{BCG15,KBG15,CLP17,KPDB17}.
At LO \nopieft contains both 2-body and 3-body contact interactions. The theory's
parameters, i.e. the 2- and 3-body low-energy constants (LECs),
were fitted to reproduce 
the $\Lambda N, NN$ scattering lengths, $^3\text{H}$ binding energy,
and the available 3,4-body $B_\Lambda$ data \cite{ConBarGal18}. 
The predictive power of the theory was tested against the measured $^5_\Lambda\text{He}$
separation energy \cite{ConBarGal18,ConBarGal19}.
The \nopieft~breakup scale can be
associated with 2-pion exchange $2m_\pi$, or the threshold value 
for exciting $\Sigma N$ pair.
These two values are remarkably close.
Assuming a typical energy scale $E_\Lambda$ of about 1 MeV, the momentum scale 
$Q\approx\sqrt{2M_{\Lambda}E_{\Lambda}}=47$~MeV/c, suggesting a \nopieft~expansion parameter $(Q/2m_{\pi})\approx 0.2$. This implies a $\nopieft$ 
LO accuracy of order $(Q/2m_{\pi})^2\approx 4$\%. 

The 3-body calculations were performed with the Stochastic Variational Method (SVM)
expanding the wave function on a correlated gaussian basis
\cite{SV95,SV98}, 
the continuum states were located using the Complex Scaling Method (CSM) 
\cite{CSM},
or the Inverse Analytic Continuation in the Coupling Constant (IACCC) Method
\cite{ch2:cit24}.  

Our main findings are: (a) The possible existence of a bound  
$\rm \Lambda nn$, or ${\rm ^3_\Lambda H^*}$ state is ruled out, 
confirming findings of previous theoretical studies 
\cite{DD59,ch1:cit5,ch1:cit6,GarFerVal07,GarVal14,ch1:cit8,ch1:cit9,BelRakSab08,AfnGib15}.
(b) The excited state of hypertriton, 
$^3_\Lambda\text{H}^*(J^\pi=3/2^+)$, is a virtual state. 
(c) The $\Lambda nn$ state is a resonance pole near the three-body threshold 
in a complex energy plane. The position of this pole
depends on the value of the $\Lambda N$ scattering length.
Within the current bounds on the $\Lambda N$ scattering length
it can either be a real resonance or a sub-threshold resonance. 


\section{Calculational details}

\subsection{Hypernuclear \nopieft at LO}
At LO the $\nopieft$ of neutrons, protons and $\Lambda$-hyperons 
is given by the Lagrangian density 
\begin{equation} 
\Lag = N^\dagger \Big(i\partial_0+\frac{\nabla^2}{2 M_N}\Big)N 
 + \Lambda^\dagger \Big(i\partial_0+\frac{\nabla^2}{2 M_\Lambda}\Big)\Lambda 
 + \Lag_{2B}+\Lag_{3B}
\label{eq:Lag} 
\end{equation} 
where $N$ and $\Lambda$ are nucleon and $\Lambda$-hyperon fields, 
respectively, and $\Lag_{2B},\Lag_{3B}$ are 2-body, and 3-body, $s$-wave 
contact interactions, with no derivatives.
These contact interactions are regularized by 
introducing a local gaussian regulator with momentum cutoff $\lambda$, 
see e.g. \cite{Bazak16}, 
\begin{equation} 
  \delta_\lambda(\rvec)=\left(\frac{\lambda}{2\sqrt{\pi}}\right)^3\,
  \exp \left(-{\frac{\lambda^2}{4}}\rvec^2\right) 
  \label{eq:gaussian} 
\end{equation} 
that smears the Dirac delta appearing in the contact terms
over distances~$\sim\lambda^{-1}$. 
This procedure yields Hamiltonian containing two-body $V_{2}$ and three-body $V_3$ interactions 
\begin{align}\label{hamilt}
 V_2 & =~\sum_{I,S} C_\lambda^{I,S} \sum_{i<j}\mathcal{P}^{I,S}_{ij} 
         \delta_{\lambda}(r_{ij})  
\cr 
 V_3 &= \sum_{I,S} D_\lambda^{I,S} \sum_{i<j<k}\mathcal{Q}^{I,S}_{ijk}\sum_{cyc}
                 \delta_{\lambda}(r_{ij})\delta_{\lambda}(r_{jk}),    
\end{align}
where $\mathcal{P}_{ij}^{I,S}$ and 
$\mathcal{Q}_{ijk}^{I,S}$ are the 2- and 3-body projection operators into
an $s$-wave isospin-spin $(I,S)$ channels. The cutoff $\lambda$ dependent parameters
$C_\lambda^{I,S}$, and $D_\lambda^{I,S}$ are the 2- and 3-body LECs, 
fixed for each $\lambda$ by the appropriate renormalization condition. 
For $\lambda$ higher than the breakup scale of the theory ($\lambda > 2m_\pi$), observables posses residual cutoff dependence, at LO $\mathcal{O}(Q/\lambda)$, 
suppressed with $\lambda$ approaching the renormalization group invariant limit $\lambda \rightarrow \infty$ \cite{ConBarGal18}.

In total there are 4 two-body ($NN$, $\Lambda N$), and 4 three-body ($NNN$, $\Lambda NN$)
LECs. The nuclear LECs $C_\lambda^{I=0,S=1}$, $C_\lambda^{I=1,S=0}$, and $D_\lambda^{I=1/2,S=1/2}$ are fitted to the deuteron binding energy, $NN$ spin-singlet scattering length $a_0^{NN}$, and to the triton binding energy, respectively. The hypernuclear two-body LECs $C_\lambda^{I=1/2,S=0}$ and $C_\lambda^{I=1/2,S=1}$ are fixed by the $\Lambda N$ <
spin-singlet $a_0^{\Lambda N}$ and spin-triplet $a_1^{\Lambda N}$ scattering lengths.
The three-body hypernuclear LECs $D_\lambda^{I=0,S=1/2}$, $D_\lambda^{I=1,S=1/2}$, and $D_\lambda^{I=0,S=3/2}$ are fitted to the experimental $\Lambda$ separation energies 
$B_\Lambda(^3_\Lambda\text{H})$, $B_\Lambda(^4_\Lambda\text{H})$, 
and the excitation energy $E_{ex}(^4_\Lambda\text{H}^*)$.

Since $a_0^{\Lambda N}$ and $a_1^{\Lambda N}$ are not well constrained by experiment, 
we consider different values both as given by direct analysis of experimental data
\cite{Alexander68}, or as predicted by several $\Lambda N$ interaction models \cite{ch2:cit3,ch2:cit4,ch2:cit5}, 
see Table~\ref{scattlengths}. For the particular values of the LECs see \cite{ConBarGal18}.

\begin{table}[htb]
\caption{Input spin-singlet $a^{\Lambda N}_0$ and spin-triplet 
  $a^{\Lambda N}_1$ scattering lengths (in fm), used to fit the 
  hypernuclear 2-body LECs. Also shown is the spin-independent combination 
  of $\Lambda N$ scattering 
  lengths $\bar{a}^{\Lambda N}=(3 a^{\Lambda N}_1+a_0^{\Lambda N})/4$.}
\begin{center}
\label{scattlengths}
\begin{tabular}{l c c c c c}
\hline \hline
 model & Reference & $a^{\Lambda N}_0$ & $a^{\Lambda N}_1$ & ${\bar a}^{\Lambda N}$\\ 
\hline
Alexander B    & \cite{Alexander68} &-1.80 &  -1.60  & -1.65 \\
NSC97f         & \cite{ch2:cit3}    &-2.60 &  -1.71  & -1.93 \\
$\chi$EFT(LO)  & \cite{ch2:cit4}    &-1.91 &  -1.23  & -1.40 \\
$\chi$EFT(NLO) & \cite{ch2:cit5}    &-2.91 &  -1.54  & -1.88 \\
\hline \hline 
\end{tabular}
\end{center}
\end{table}

\subsection{The Stochastic Variational Method}
The $A$-body Schr\"{o}dinger equation is solved expanding the wave function $\Psi$ 
in correlated gaussians basis \cite{SV95}
\begin{equation} \label{corrgauss}
   \Psi=\sum_i c_i~\psi_i 
       = \sum_i c_i~\hat{\mathcal{A}}
         \left\{{\rm exp}\left(-\frac{1}{2} {\bf x}^T A_i {\bf x}\right) 
                \chi^i_{S M_S} \xi^i_{I M_I} \right\},
\end{equation}
where $\hat{\mathcal{A}}$ stands for the antisymmetrization operator over nucleons, ${\bf x}=({\bf x}_1, ..., {\bf x}_{A-1})$ denotes a set of Jacobi vectors, and $\chi^i_{S M_S}$ ($\xi^i_{I M_I}$) is the spin (isospin) part. The information about interparticle correlations is contained in the $(A-1)$ dimensional positive-definite symmetric matrix $A_i$. Once we fix all basis functions $\psi_i$, both energies 
and coefficients $c_i$ are obtained through diagonalization of the Hamiltonian matrix. The $A(A-1)/2$ nonlinear variational parameters contained in each $A_i$ matrix are determined using the Stochastic Variational Method (SVM) \cite{SV95,SV98}.

Unlike bound states, continuum wave functions are not square-integrable. 
Therefore, resonances or virtual states can not be directly described 
using an $L^2$ basis set of correlated gaussians. Techniques such as CSM or IACCC have 
to be used to study such states with a correlated gaussians. 
Below we discuss in some detail the techniques we applied in our study.

\subsection{The Complex Scaling Method}

The CSM \cite{CSM} is a reliable tool to study few-body resonances \cite{AoyMyoKatIke06}. The basic idea in the CSM is to locate resonances introducing
complex rotation of coordinates and momenta 
\begin{equation}\label{csm}
  U(\theta){\bf r} = {\bf r}e^{{\rm i}\theta},~~~~U(\theta){\bf k} = {\bf k}e^{-{\rm i}\theta},
\end{equation}
that transforms the continuum states into integrable $L^2$ states. This transformation
rotate continuum state energies by $2\theta$ uncovering a section of the second energy
plane between the real axis and a ray defined by $|\text{arg} E| = 2\theta$,
exposing resonances with argument 
$\theta_r={\arctan}({\Gamma}/{2 E_r})/2$
smaller than $\theta$.
Using gaussian regulator (\ref{hamilt}) the rotation angle is restricted
to be $\theta < \frac{\pi}{4}$, to prevent divergence of the rotated gaussian,
limiting the scope of the CSM. 

The SVM method use the variational principle as a tool to optimize the nonlinear
basis parameters $A_i$ \eqref{corrgauss}, minimizing the basis size. 
This do not apply
to resonance states, making it a highly non trivial problem to choose the 
appropriate basis.
Here, we present a new efficient procedure to determine the basis set for an accurate description of resonance states. 
To optimize the basis, we supplement the Hamiltonian $H$ 
with an additional harmonic oscillator (HO) trap
\begin{equation}
  H^{trap}(b)=H+V^{HO}(b),~~~V^{HO}(b)=\frac{\hbar^2}{2mb^4} \sum_{j<k}r_{jk}^2,
\end{equation}
where  $m$ is an arbitrary mass scale, and $b$ is the HO trap length. The potential $V^{HO}(b)$ gives rise to a HO spectrum of the ground and excited states which is affected by the presence of a resonance in the Hamiltonian $H$ \cite{ch2:cit17}. For a given trap length $b$ we select basis states $\psi_i$ \eqref{corrgauss} using the SVM, 
optimizing the variational parameters for the ground state energy and then subsequently for excited states energies up to $E_{max}>E_r+\Gamma/2$. 
The SVM procedure prefers basis states which promote interparticle distances $r_{jk}$ in a specific region given by the trap length $b$.
Increasing $b$ we enlarge the typical radius of the correlated gaussians $\psi_i$. 
For large enough $b$, the CSM resonance solution for the Hamiltonian $H$ starts to stabilize and both the short range and the suppressed long range asymptotic parts of a resonance wave function are described sufficiently well. 
In order to further enhance the accuracy of our CSM solution, we use a grid $\{b_k\}$, 
of a HO trap lengths, and for each grid point we independently select 
correlated gaussians basis. Then we merge basis states determined for each $b_k$ into a larger basis while ensuring linear independence and numerical stability of the overlap 
matrix.
We have found that this procedure works well for both narrow,
and broad resonances. 
\subsection{Inverse Analytic Continuation in the Coupling Constant Method}
The Analytic Continuation in the Coupling Constant (ACCC) method \cite{ch2:cit18} has been successfully applied in various calculations of few-body resonances and virtual states \cite{ch2:cit19,ch2:cit20}. Moreover, it was pointed out that the ACCC method provides rather convenient way how to extend applicability of the SVM into the continuum region \cite{ch2:cit19,ch2:cit21}. We consider a few-body Hamiltonian consisting of the
physical part $H$ and an auxiliary attractive potential $V^{aux}$ 
\be \label{accchamilt}
   H^{IACCC}=H+\alpha~V^{aux},
\ee
which introduces a bound state for a certain value of $\alpha$, but ensures that the physical dissociation thresholds for the various subsystems remain unaffected.
By decreasing the strength $\alpha$ the bound state moves closer to the threshold and for a certain $\alpha_0$ it turns into a resonance or virtual state. It has been demonstrated 
for a two-body system that in the vicinity of the branching point $\alpha_0$ the square 
root of an energy $k=\sqrt{E}$ behaves as $k\approx(\alpha-\alpha_0)$ for $s$-wave ($l=0$) and $k\approx\sqrt{\alpha-\alpha_0}$ for $l>0$ \cite{ch2:cit18}. Defining new variable $x=\sqrt{\alpha-\alpha_0}$ one obtains two branches $k(x)$ and $k(-x)$ where the former one describes motion of the S-matrix pole assigned to a bound state on a positive imaginary $k$-axis to the third quadrant of a $k$-plane. Using analyticity of the function $k(x)$ one can continue from a bound region $\alpha>\alpha_0$ to a resonance region $\alpha<\alpha_0$. In practice this is done by constructing a Pad\'{e} approximant 
\begin{equation}
k(x)\approx {\rm i}\frac{\sum_{j=0}^M c_j x^j}{1+\sum_{j=1}^N d_j x^j}
\label{approximant}
\end{equation}
for the function $k(x)$ using $M+N+1$ bound state solutions $\{(x_j,k_j); j=1,\dots,M+N+1\}$ for different values of $\alpha>\alpha_0$. The evaluation of the Pad\'{e} approximant (\ref{approximant}) at $x=\sqrt{-\alpha_0}$ yields complex $k$ which is assigned to the physical resonance solution $k^2=E_r-{\rm i}\Gamma/2$ corresponding to the Hamiltonian $H$. For more details regarding the ACCC method see \cite{ch2:cit22}.

The ACCC method suffers from two drawbacks which are predominantly of numerical nature. The first issue is high sensitivity of the numerical solution to precise determination of the branching point value $\alpha_0$ \cite{ch2:cit18}. The second obstacle appears with increasing orders $M$ and $N$ of the Pad\'{e} approximant (\ref{approximant})
when the numerical solution starts to deteriorate. 

Rather recently Hor\'{a}\v{c}ek et al. \cite{ch2:cit24} have introduced a modified version of the ACCC method called the Inverse Analytic Continuation in the Coupling Constant (IACCC) method which provides more robust numerical stability. Starting in the same manner as in the ACCC case, we consider the Hamiltonian (\ref{accchamilt}) and calculate series of bound states for different values of $\alpha >\alpha_0$. Next, we construct a Pad\'{e} approximant of a function $\alpha(\kappa)$, where $\kappa=-{\rm i}k$, using a relevant set of bound state solutions
\begin{equation} \label{iacccpade}
   \alpha(\kappa) \approx \frac{P_M(\kappa)}{Q_N(\kappa)}
               = \frac{\sum_{j=0}^M c_j \kappa^j}{1+\sum_{j=1}^N d_j \kappa^j}.
\end{equation}
The parameters of the physical resonance or virtual state pole are then 
readily obtained by setting $\alpha=0$ as the physical root of a simple polynomial 
equation $P_M(\kappa)=0$.

To ensure that the properties of the 2-body part of the Hamiltonian, such as scattering 
lengths or deuteron binding energy, remain unaffected, we choose the auxiliary
potential to be an attractive 3-body force. The natural choice
is to select it to have the same form as the \nopieft 3-body potential
~\eqref{hamilt},
\be \label{v3iacc}
   V^{IACCC}_3 = d_\lambda^{I,S} \sum_{i<j<k}\mathcal{Q}^{I,S}_{ijk}\sum_{cyc}
   {e}^{-\frac{\lambda^2}{4}\left(r_{ij}^2+r_{jk}^2\right)},
\ee
where the amplitude $d_\lambda^{I,S}$ defines its strength, corresponding to the parameter $\alpha$ in Eq. \eqref{accchamilt}, and is negative for an attractive auxiliary potential. 

The accuracy of our IACCC resonance solutions in the fourth quadrant of the complex 
energy plane, ${\rm Re}(E)>0$, ${\rm Im}(E)<0$, are better then 
$\approx 10^{-3}~{\rm MeV}$. These results compare very well with the CSM calculations
in their region of applicability $\theta < \pi/4$. 

\section{Results}

Using \nopieft at LO with the LECs fitted to the available data as described earlier
\cite{ConBarGal18}, we find no bound $\Lambda n n$ or $^3_\Lambda\text{H}^*$ states.
Further examining the hypothetical existence of these
states, we found that they are incompatible 
with the well measured $A=4,5$ hypernuclear spectrum.

As we have already pointed out, the possible existence of bound 
$\Lambda nn$ and $^3_\Lambda\text{H}^*$ states has been quite convincingly ruled out in several theoretical studies 
\cite{DD59,ch1:cit5,ch1:cit8,GarFerVal07,GarVal14,ch1:cit9,ch1:cit6}. 
Our \nopieft findings support their conclusions. 

\subsection{A $\Lambda nn$ resonance?}

\begin{figure}[t!]
\includegraphics[width=0.5\textwidth]{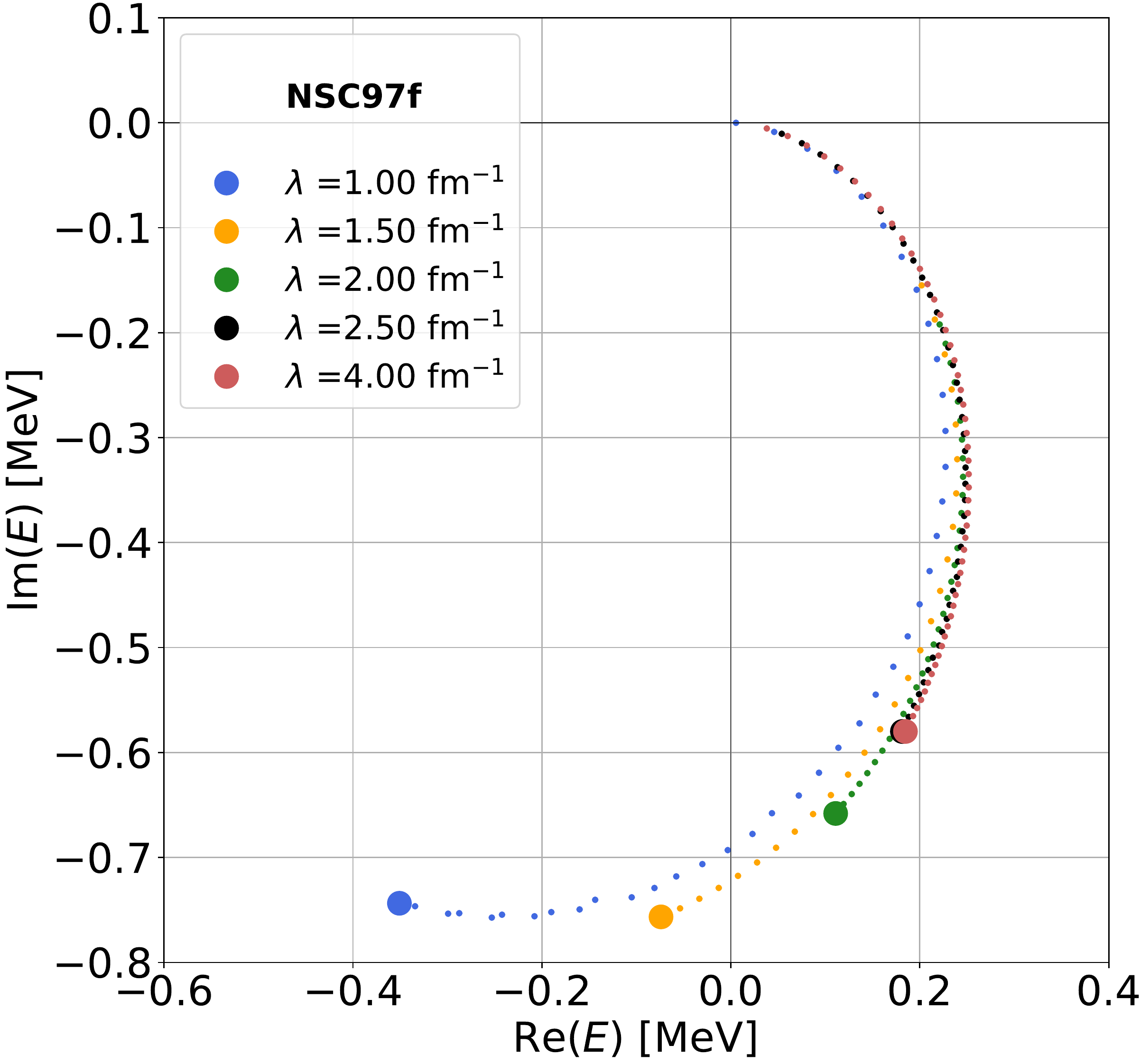}
\caption{\label{lnn_cutoff} 
Trajectories of the $\rm \Lambda nn$ resonance pole in the complex energy plane 
determined by a decreasing attractive strength of the auxiliary three-body force $d_\lambda^{I=1,S=1/2}$ for several cutoffs $\lambda$ and the NSC97f set of
$\Lambda N$ scattering lengths. Small dots mark IACCC solutions for different $d_\lambda^{I=1,S=1/2}$, larger symbols stand for the physical position of the 
$\Lambda nn$ pole ($d_\lambda^{I=1,S=1/2}=0$). 
Notice the almost overlapping trajectories for $\lambda=2.50~{\rm fm^{-1}}$ and $\lambda=4.00~{\rm fm^{-1}}$.}
\end{figure}

We start our study of three-body hypernuclear continuum states with the $\Lambda nn$
system. 
To understand the cutoff dependence of our theory 
we present, in Fig.~\ref{lnn_cutoff}, the trajectories 
$E_{\Lambda nn}(d_\lambda^{I=1,S=1/2},\lambda)$ of the $\Lambda nn$ resonance pole, 
calculated using the IACCC method for different values of cutoff $\lambda$, 
and for a representative set of $a^{\Lambda N}_s$ - NSC97f.
With decreasing attraction of $V_{3}^{IACCC}$, the resonance poles move along 
a circular trajectory in the complex energy plane starting from the $\Lambda+n+n$ threshold to the physical end point where $d_\lambda^{I=1,S=1/2}=0$. 
From the figure it can be seen that the 
the trajectories $E_{\rm \Lambda nn}(d_\lambda^{I=1,S=1/2},\lambda)$
and the physical end points 
converge with increasing cutoff, and already at
$\lambda=2.5~{\rm fm^{-1}}$ we reach stabilized results. 

Repeating the same calculations for all sets of scattering lengths given
in Table~\ref{scattlengths}, we find that
regardless the cutoff value, the imaginary part of the physical solution ${\rm Im}(E^\lambda_{\rm \Lambda nn})$ lies in the interval $-1.32\leq{\rm Im}(E^\lambda_{\rm \Lambda nn})\leq -0.58$ MeV for all $a^{\Lambda N}_s$ sets. 
In contrast, the real part ${\rm Re}(E^\lambda_{\rm \Lambda nn})$ exhibit large cutoff dependence. 
As shown in Fig.~\ref{lnn_cutoff} for the NSC97f case, the pole moves with increasing 
$\lambda$ from the unphysical part of the Riemann sheet 
(${\rm Re}(E)<0$, ${\rm Im}(E)<0$; third quadrant) towards the physical one (${\rm Re}(E)>0$, ${\rm Im}(E)<0$; fourth quadrant). 

\begin{figure}[h!]
\includegraphics[width=0.5\textwidth]{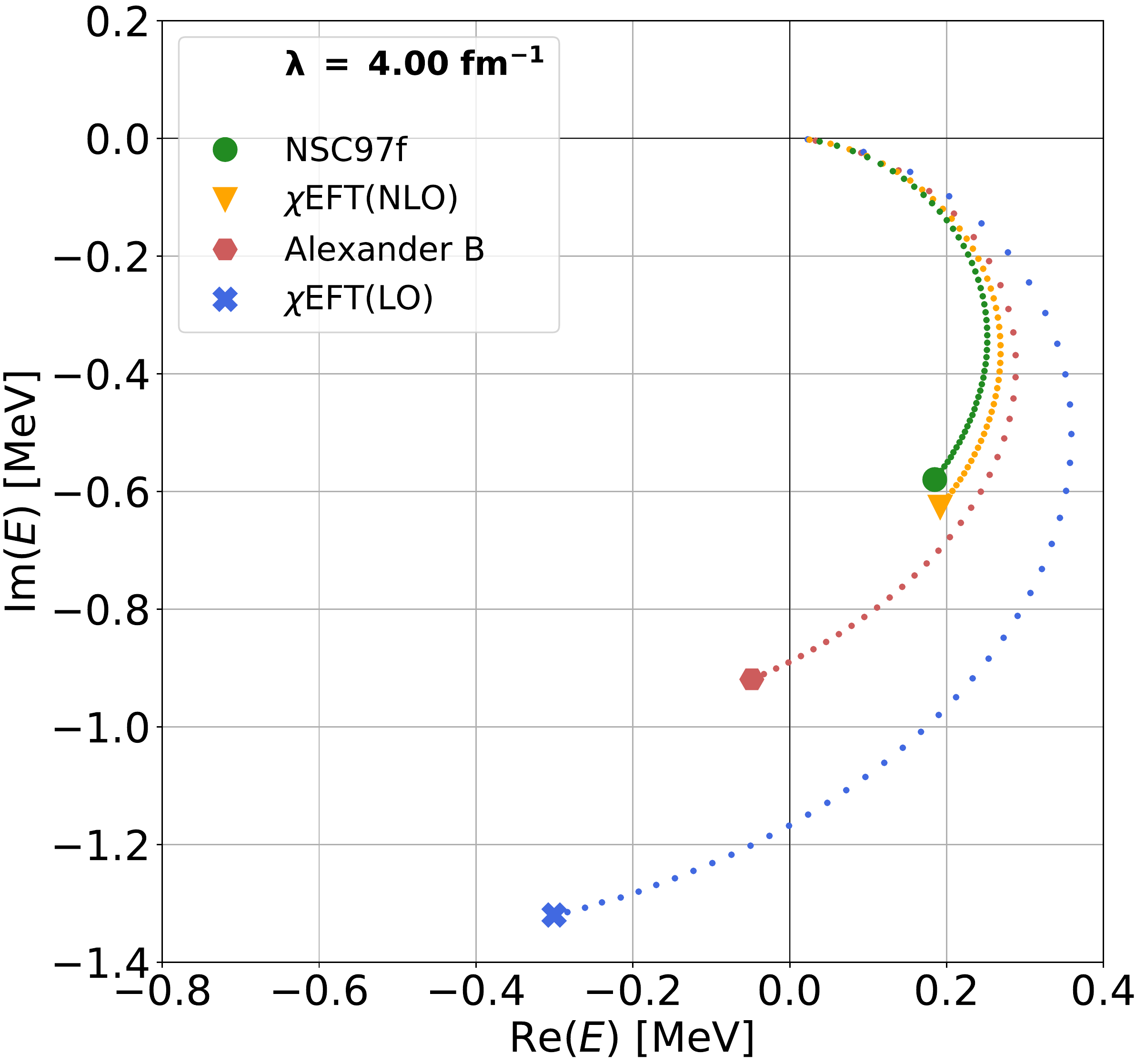}
\caption{\label{lnn_models}Trajectories of the $\rm \Lambda nn$ resonance pole in the complex energy plane determined by a decreasing attractive strength $d_\lambda^{I=1,S=1/2}$ for selected sets of $\Lambda N$ scattering length, calculated 
at $\lambda=4.00~{\rm fm^{-1}}$. Larger symbols stand for the physical position of the 
$\Lambda nn$ pole ($d_\lambda^{I=1,S=1/2}=0$). 
}
\end{figure}

In Fig.~\ref{lnn_models} we compare the trajectories 
$E_{\rm \Lambda nn}(d_\lambda^{I=1,S=1/2},\lambda)$ 
for the different values of $\Lambda N$ scattering lengths, Table~\ref{scattlengths}, 
at cutoff $\lambda=4~{\rm fm^{-1}}$.
From the figure, we can deduce that the existence of a physically observable 
$\Lambda nn$ resonance is very sensitive to the $\Lambda N$ interaction. 
The latter must be strong
enough to ensure the pole's location in the fourth quadrant of a complex energy plane.
The figure and Table~\ref{scattlengths} show that with increasing size of the spin-averaged scattering length 
$\bar{a}^{\Lambda N}={3}/{4}a^{\Lambda N}_1+{1}/{4}a^{\Lambda N}_0$  the 
$\Lambda nn$ pole trajectories move closer to the $\Lambda +n+n$ threshold. 
Moreover, by increasing the cutoff $\lambda$ the physical $\Lambda nn$ pole is shifted closer to or into the fourth quadrant.
In this sense the pole position in the renormalization group invariant limit
 $\lambda \rightarrow \infty$ could be considered as the most favorable to the existence of an observable resonance. 
Nevertheless, in the $\lambda \to \infty$ limit only two sets of 
$a^{\Lambda N}_s$ - NSC97f and $\chi$EFT(NLO) undoubtedly predict a physical resonance. 
From the results shown in Fig.\ref{lnn_models} we can roughly estimate that  
$\bar{a}^{\Lambda N} \approx 1.7~{\rm fm^{-1}}$ is the minimal value for
the $\Lambda nn$ pole to enter the fourth quadrant, becoming a physical resonance.
It should be noted that though
the size of $\bar{a}^{\Lambda N}$ plays a dominant role,
one should take into account also the effect of the three-body force which might 
introduce more complicated dependence on $a_0^{\Lambda N}$ and $a_1^{\Lambda N}$.

\subsection{The hypertriton excited state $^3_\Lambda\text{H}^*(J^\pi=3/2^+)$}
\begin{figure}[t!]
\includegraphics[width=0.5\textwidth]{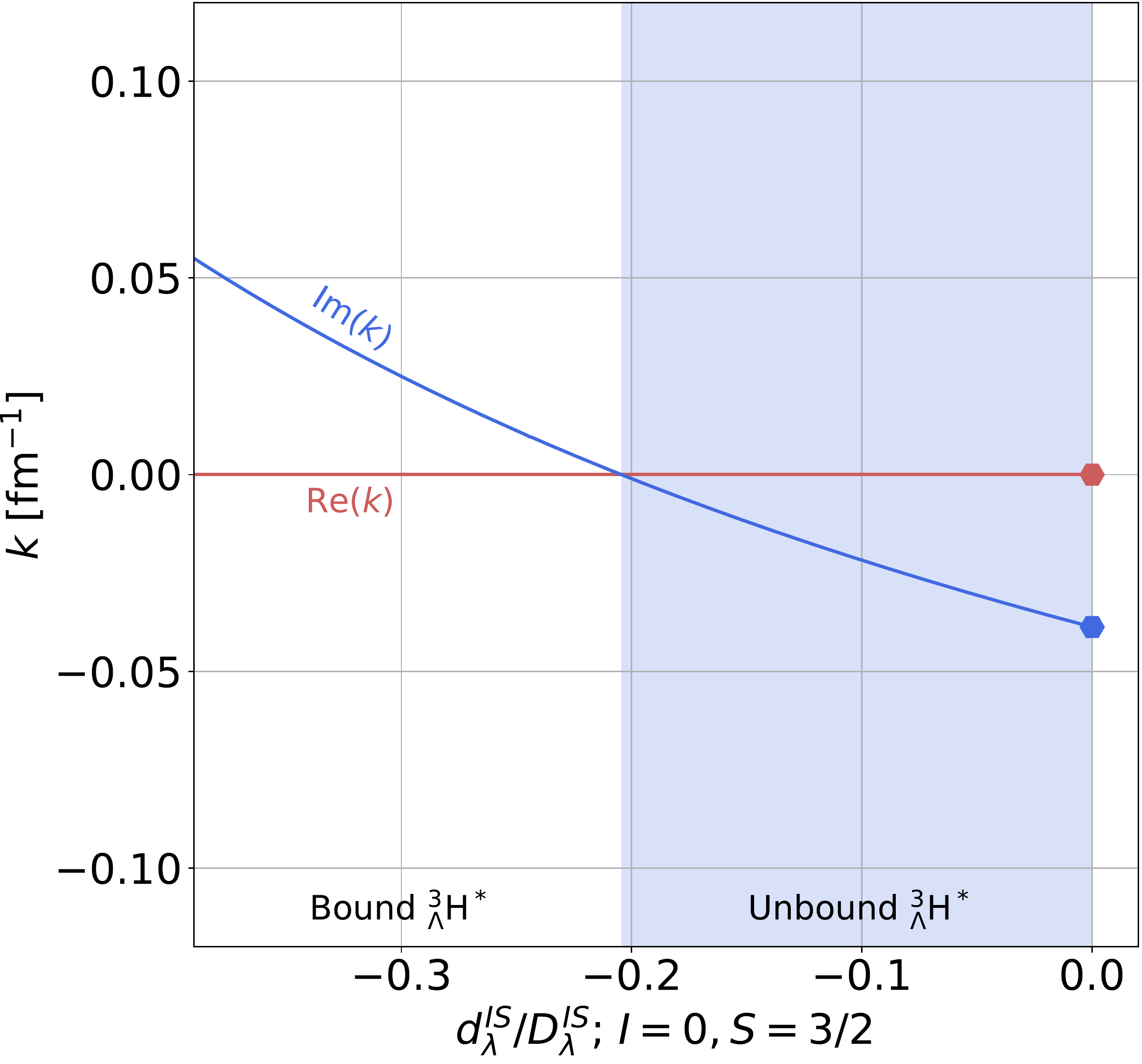}
\caption{\label{fig5} Imaginary (blue) and real (red) parts of the $\rm ^3_\Lambda H^*$ pole momentum $k$ as a function of $d_\lambda^{I=0,S=3/2}$, normalized to the physical three-body LEC $D_\lambda^{I=0,S=3/2}$. Unbound region is determined through the IACCC method. Dots mark the physical solution with for $d_\lambda^{I=0,S=3/2}=0$.
}
\end{figure}

The excited state of the hypertriton $^3_\Lambda\text{H}^*(J^\pi=3/2^+)$ might be considered as a good candidate for a near-threshold resonance. Indeed, several works demonstrated an emergence of a bound state by increasing rather moderately the $\Lambda N$ interaction strength. Applying the IACCC method we follow the pole trajectory given by the amplitude of auxiliary 3-body force $d_\lambda^{I=0,S=3/2}$ from a bound region to its physical position in a $\Lambda$+deuteron ($\Lambda+d$) continuum. In Fig.~\ref{fig5} we show the $\rm ^3_\Lambda H^*$ pole momentum 
$k=\sqrt{2\mu_{\Lambda d} [E({\rm ^3_\Lambda H^*})-E_B({\rm ^2H})]}$, 
$\mu_{\Lambda d}=m_{d}m_\Lambda/(m_{d}+m_\Lambda)$, 
as a function of $d_\lambda^{I=0,S=3/2}$ for Alexander B $\Lambda N$ scattering lengths and $\lambda=6~{\rm fm^{-1}}$. We observe that with a decreasing auxiliary attraction the imaginary part of the momentum Im($k$) decreases from positive value (bound state) 
to a negative value (unbound state) whereas the real part Re($k$) remains equal to zero. This behavior is regarded as definition of a virtual state \cite{ch3:cit12}.

Repeating the calculations for various cutoffs and different ${\Lambda N}$ scattering 
lengths, Table \ref{scattlengths}, we find $^3_\Lambda\text{H}^*$
to be a virtual state in all considered cases. 
As we have seen in the $\Lambda nn$ calculations, the energy of the virtual state 
$E_v$ is stabilized at cutoffs $\lambda \geq 4~{\rm fm^{-1}}$.

\begin{figure}[h!]
\includegraphics[width=0.5\textwidth]{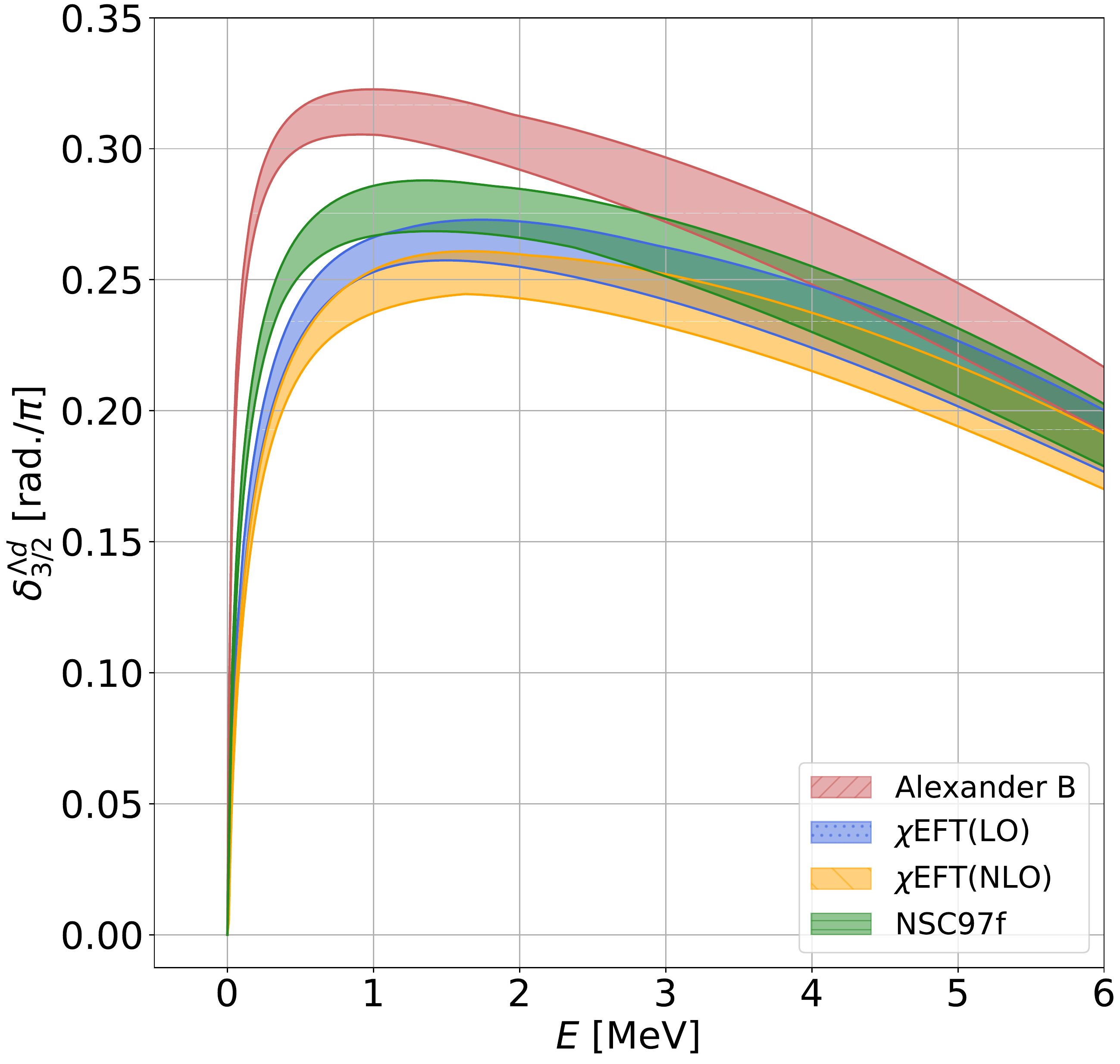}
\caption{\label{ld-phaseshifts} $S$-wave $\Lambda d$ phase-shifts in 
the $J^\pi=3/2^+$ channel $\delta^{\Lambda d}_{3/2}$ as a function of energy 
$E$ above the $\Lambda+d$ threshold, extracted from the continuum level density 
of the rotated CSM spectra. The phase-shifts are calculated for cut-off 
$\lambda = 6~{\rm fm^{-1}}$ and several $\Lambda N$ interaction strengths. 
Shaded areas mark uncertainty introduced by the rotation angle $\theta$ 
within interval $15^{\circ}<\theta<20^{\circ}$. 
}
\end{figure}

The existence of the $\rm ^3_\Lambda H^*$ virtual state is further confirmed by the CSM. 
We do not see any sign of resonance for all sets of ${\Lambda N}$ scattering lengths,
cutoffs, or auxiliary 3-body force values $d_\lambda^{I=0,S=3/2}$. Odsuren et al. \cite{Odsuren17} have showed that the rotated discretized CSM continuum spectra 
reflect phenomena such as near-threshold virtual states, although one would naively
assume that virtual states having $|\text{arg} E|=\pi/2$ are beyond the
reach of the CSM. From continuum level density they have extracted the
scattering phase shifts which revealed enhancement due to the vicinity of a the pole
 \cite{Odsuren17,Odsuren14}. 
Following this approach we calculated the $\Lambda d$ $s$-wave phase shifts
$\delta_{3/2}^{\Lambda d}$ for the $J=3/2$ channel.
The calculated phase shifts, presented in Fig.~\ref{ld-phaseshifts}, exhibit clear enhancement close to threshold implying proximity of a pole. 
The shaded areas in the figure reflect the phase shift dependence on 
rotation angle $\theta$, which we checked for a rather broad interval $15^{\circ}<\theta<20^{\circ}$.

The scattering length $a^{\Lambda d}_{3/2}$ and effective range 
$r^{\Lambda d}_{3/2}$ extracted from the $\Lambda d$ phase shifts 
reveal through their sign, negative $a^{\Lambda d}_{3/2}$ and positive $r^{\Lambda d}_{3/2}$,
the existence of a virtual state \cite{ch3:cit14}. 
Using $a^{\Lambda d}_{3/2}, r^{\Lambda d}_{3/2}$ the virtual state binding momentum $k_v=\sqrt{2\mu_{\Lambda d}E_v}$ can be approximated by
\begin{equation} \label{ere-pole}
   k_v=\frac{i}{r^{\Lambda d}_{3/2}}
               \left(1-\sqrt{1-\frac{2~r^{\Lambda d}_{3/2}}{a^{\Lambda d}_{3/2}}}\,\right).
\end{equation}

In Table~\ref{lambdad} we present
the IACCC results for $E_v$, and an estimate 
$a^{\Lambda d}_{3/2}=-i/\sqrt{2\mu_{\Lambda d} E_v}$ for 
the scattering length, together with
the scattering parameters $a^{\Lambda d}_{3/2}$ and $r^{\Lambda d}_{3/2}$
extracted from the CSM calculations and the resulting estimate for $E_v$, 
Eq. \eqref{ere-pole}. 
Inspecting the table, one might naively expect clear monotonic dependence of $E_v$ on 
the spin-triplet scattering length $a_1^{\Lambda N}$. However, the dominance of 
$a_1^{\Lambda N}$ is undermined by the 3-body force in the $(I,S)=(0,3/2)$ channel, 
fixed by $B_\Lambda(^4_\Lambda\text{H}^*)$.
Comparing the IACCC and CSM results, one clearly see that both approaches are in mutual
agreement, they exhibit the same dependence on the $\Lambda N$ interaction strength, though, the CSM yields larger estimates for $|E_v|$.
It is a well known drawback of the CSM that eigenvalues in a vicinity of the threshold start to be affected by inaccuracies caused by complex arithmetic.

\begin{table}[t!]
\caption{\label{lambdad} Calculated $\Lambda d$ scattering lengths $a^{\Lambda d}_{3/2}$, effective ranges $r^{\Lambda d}_{3/2}$, and virtual state energies $E_v$ in 
$J=3/2$ channel for several $\Lambda N$ interaction strengths and cutoff $\lambda=6~{\rm fm^{-1}}$. Results of two different methods are presented - the continuum level 
density of rotated CSM spectra and the IACCC method. 
For the CSM we obtain $E_v$ using relation \eqref{ere-pole}, for the IACCC using the 
relation $a^{\Lambda d}_{3/2}=-i/\sqrt{2\mu_{\Lambda d} E_v}$. The scattering length and effective range 
are given in fm, $E_v$ in MeV.}
\begin{tabular}{lccccc}
\hline\hline
     &  CSM  &  &  & IACCC & \\ 
     & $a^{\Lambda d}_{3/2}$ & $r^{\Lambda d}_{3/2}$ & $E_v$ & $a^{\Lambda d}_{3/2}$ 
     & $E_v$ \\ 
\hline
Alexander B    & -17.3 & 3.6 & -0.08 & -25.7 & -0.042 \\
NSC97f         & -10.8 & 3.8 & -0.18 & -16.1 & -0.108 \\
$\chi$EFT(LO)  & -8.5  & 3.5 & -0.28 & -12.8 & -0.169 \\
$\chi$EFT(NLO) & -7.6  & 3.6 & -0.34 & -11.7 & -0.205 \\
\hline\hline
\end{tabular}
\end{table}

Concluding this section, we see that at LO \nopieft firmly predicts the excited state of hypertriton ${\rm ^3_\Lambda H^*}(J^\pi=3/2^+)$ to be a virtual state 
in the vicinity of the $\Lambda-d$ threshold. 
This result has important implications for prospective experimental search of this 
state. Experimental observation of ${\rm ^3_\Lambda H^* }$ as a resonance state seems to be
highly unlikely. Instead, there is a near-threshold virtual state which should be seen through the enhancement of $s$-wave $\Lambda d$ phase shifts in the $J=3/2$ channel as demonstrated in Fig.~\ref{ld-phaseshifts}.  

\section{Conclusions}

In this work we have presented the first comprehensive \nopieft study of continuum 
hypernuclear $\Lambda NN$ trios.
The underlying nucleon and hyperon interactions were described within a \nopieft at LO,
with the LECs fixed by 2-body low energy observables and experimental input from 3- and 4-body $s$-shell systems. 
The $\Lambda nn$ and ${\rm ^3_\Lambda H^*}$ energies were then obtained as 
predictions of the theory. 
In view of poor low energy $\Lambda N$ scattering data we considered several sets of $\Lambda N$ scattering lengths, whereas the $NN$ interaction remained constrained by experiment \cite{ConBarGal18}.

Few-body wave functions were described within a correlated gaussians basis. 
Bound state solutions were obtained using the SVM. The continuum region
was studied employing two independent methods - the IACCC method and CSM. 

The \nopieft predicts that both $\Lambda nn$ and ${\rm ^3_\Lambda H^*}$ 
are unbound. Tuning the 3-body LECs to put the $\Lambda nn$ 
or ${\rm ^3_\Lambda H^*}$ binding energy on threshold, yielded considerable
discrepancy between the calculated and measured $B_\Lambda$ in
the $A=4,5$ hypernuclei. 
Our findings further strengthen the conclusions of previous theoretical studies
that both states are unbound \cite{DD59,ch1:cit5,ch1:cit8,GarFerVal07,GarVal14,BelRakSab08,ch1:cit9,ch1:cit6,AfnGib15}.   

Our LO \nopieft calculations predict $\Lambda nn$ and ${\rm ^3_\Lambda H^*}$ 
to be near-threshold continuum states. 
We thus anticipate that the EFT truncation error is small due to low 
characteristic momenta and thus higher order corrections would not change our 
results qualitatively.
We conclude that position of the $\Lambda nn$ pole
depends strongly on the spin independent scattering length $\bar{a}^{\Lambda N}$. 
For $\bar{a}^{\Lambda N} \geq 1.7~{\rm fm^{-1}}$ the $\Lambda nn$ pole becomes 
a physical resonance close to threshold with $E_r \leq 0.3~{\rm MeV}$, and a large width 
most likely in the range $1.16 \leq \Gamma \leq 2.00~{\rm MeV}$. 
If observed, the position of the $\Lambda nn$ resonance can yield tight constraints
on the $\Lambda N$ scattering length.
We note, owever, that the exact position of the $\Lambda nn$ depends on  
both on $a_0^{\Lambda N}$ and $a_1^{\Lambda N}$, and also on subleading \nopieft terms neglected
here. 
The excited state of hypertriton ${\rm ^3_\Lambda H^*}$ was firmly predicted to be a 
near-threshold virtual state regardless of the value of $a^{\Lambda N}_s$.
We have demonstrated that this virtual state has a strong effect on
the $\Lambda d$ $s$-wave  phase shifts in $J^\pi=3/2^+$ channel.  

\section*{Acknowledgments}
We are grateful to Avraham Gal for valuable discussions and careful reading of the manuscript. 
This work was partly supported by the Czech Science Foundation GACR grant 19-19640S,
The work of NB was supported by the Pazy Foundation and by the 
Israel Science Foundation grant 1308/16.

\end{document}